\documentclass[a4,50pt]{amsart}%
\usepackage{amssymb}
\usepackage{pdflscape}
\usepackage{amsmath}
\usepackage{mathrsfs}
\usepackage{amsfonts}
\usepackage{graphicx}%
\usepackage[all]{xy}
\theoremstyle{plain}

\newtheorem*{remark}{Remark}

\numberwithin{equation}{section}
\begin{document}
%\begin{landscape}
\title[Physical Roots of QSC]{The Generation of Quantum Hamiltonian Evolution from A Pseudo-Measurement Process}
\author{M\ F\ Brown}
\maketitle

\begin{abstract}
Here we shall consider the idea that the Hamiltonian evolution of a quantum system is generated by sequential observations of the system by a `pseudo-apparatus'.  This representation of Hamiltonian dynamics, originally discovered by Belavkin, is a canonical dilation of the Schr\"odinger equation that reveals a Minkowski space structure outside of the context of Special Relativity. In particular, this formalism gives rise to the notion of a \emph{Boosted Schr\"odinger equation} referred to a dilated time increment which is the manifestation of a Lorentz transform in the Minkowski space of the apparatus.
\end{abstract}
\tableofcontents
\section{Introduction}
Here we consider a quantum system referred to as a quantum object. This object is under observation by a pseudo measurement apparatus whose purpose is to simply detect the quantum object without measuring any potential state that the object is supposed to be in. This detection is derived as \emph{the process by which an evolution of the object is generated}. It is by virtue of this measurement apparatus that time is assigned to the object, and the energy spectrum associated with the object arises from the interaction corresponding to a measurement.

\subsection{Pseudo-Apparatus}
The pseudo-apparatus is a system having two degrees of freedom, and a transition from one to the other is used to infer that a change has taken place in the quantum system under observation. The terminology \emph{pseudo} is used as the apparatus is represented in a pseudo-Hilbert space, that is remarkably derived having a Minkowski metric; some of the details of this discovery, by Belavkin, we shall see in due coarse.

The process of pseudo-measurement corresponds to a pseudo-unitary interaction of the pseudo-apparatus with the quantum object; this is regarded as an observation. The term pseudo-unitary refers to the fact that in a pseudo-Hilbert space the unitarity of any interaction must be given with respect to the pseudo-involution induced by the pseudo-metric, which in this case is the Minkowski metric.  We shall see as we progress that  the Hamiltonian of the quantum object appears in the process of interaction, and in the absence of the interaction the quantum system is considered not to evolve. This means that the evolution of the quantum system is actually stimulated by the measurement process. The term `pseudo' will also be dropped in many cases henceforth, but this mathematical technicality should not be forgotten.

The two degrees of freedom of the apparatus may be unambiguously identified with \emph{future} and \emph{past}, or respectively \emph{input} and \emph{output}, and it is the measurements made by this apparatus that generate a flow of time. The apparatus is initially considered to be prepared in a state of future/input and on interacting with the quantum system the apparatus generates a past/output. Generally speaking, observable events are in the future and only become elements of the past once they are observed. In this way `future' may simply be understood as the domain of incoming information and `past' the domain of outgoing information \cite{Be00b}.

What we find is that this (pseudo) apparatus may be regarded as an apparatus intrinsic to all evolution, for it establishes a method by which evolution may be generated by observations. It will be outlined here that such observations, corresponding to the interaction of the quantum system with the apparatus, are formulated precisely as differential increments in the system's evolution, but the reader is referred to \cite{MFB11} for more details and to \cite{Be92b} for the general theory.

\subsection{Hamiltonian Dynamics from Quantum Filtering}
The fact that the evolution of a quantum system may be derived from this so called \emph{algebraic realization of calculus} (that is the interaction of the object with this intrinsic pseudo-apparatus) is one thing, but this formalism also realizes the instants of the measurement as massless particles. This ultimately gives rise to a quantum object dynamics that is generated by spontaneous interactions with a massless field, and these interactions are precisely the differential increments that generate the evolution of the quantum system.

As one might expect from a formalism with massless particles in a Minkowski space there are Lorentz transformations appearing in this theory. Their role is to increase, or decrease, the interaction intensity of the quantum system with the massless particles. We shall see that the effect that the Lorentz transforms have on the interaction operator is to `boost' the time increment in the outgoing Schr\"odinger equation.

As quantum Hamiltonian evolution is to be generated by a sequence of random measurements it is appropriate to discuss the role of waves and particles in this setup. Generally speaking, potential events should be described primarily by waves and  particles are simply realized as actual events arising from the process of measurement. This is in accordance with \cite{Be00b} and also corresponds to the view of Lawrence Bragg. It amounts to the statement that the future is a wave input and the past is a particle output. However, this outlook also admits that a random set of points in the future may be considered to correspond  to eventual interactions i.e. a random set of future points that have been considered for the purposes of mathematics. In this way any such point in the future may be regarded as %the location of a wavefront, and such wavefronts may also be referred to as `particles' in an appropriate context.
a particle, although such future-particles should only be understood as potential interactions. This convention is also rigorous for the purposes of physics as the Fock-space formulation of stochastic noise has cause for referring to `particles of noise'.

In the case of standard quantum filtering, see for example \cite{Be92,Be00b}, the conditioning of the measurement dynamics, with respect to the measurements, describes the process of inferring pure states of the quantum system under observation. Here things are a little different because tracing over the apparatus does not give rise to a mixture corresponding to von Neumann's projection postulate. Instead, the quantum system state, $\varrho$ say, is mapped to the derivative $\dot{\varrho}:=\mathrm{i}[\varrho,H]$ so that one may understand a pseudo-decoherence of the quantum state as it is transformed into its derivative; note that $\verb"Tr"[\dot{\varrho}]=0$, characterizing a pseudo-state. % and the second is that the entropy of the null state of the compound object-particle system is zero. %although one may question the meaning of entropy here. Also note: pure states have no entropy!

In this measurement process we do not have filtering in the sense that we are making a prediction of a particular quantum state by the filtering of a decoherent mixture of such states. However, we are predicting that a quantum system simply exists by virtue of this measurement process. This resembles a trivial filtering process if one trivially replaces the pseudo-decoherence $\varrho\mapsto\dot{\varrho}$ with the map $\varrho\mapsto\dot{\varrho}+o$, where the latter introduces $o$ to explicitly represent the fact that there could be nothing there. Thus the filtering now corresponds to a sequential act of differentiation $\varrho\mapsto\dot{\varrho}\mapsto\ddot{\varrho}\mapsto\cdots$
and the Hamiltonian dynamics of a quantum state may be interpreted as a quantum filtering of the system by the interactions with massless particles. This gives us dynamics in which a massless incoming wave from the future generates an outgoing Hamiltonian dynamics in the past.

This dilation of Hamiltonian evolution comes with a very clear and intuitive picture, although seemingly counterintuitive at first. What we find is that the description of a system ultimately results from the occurrence of a  boundary between  the future and the past. On the one hand, we have a system that may at first be simply identified with the boundary. Whilst on the other hand we have a wave (identified with the apparatus) propagating from the future into the past, and such is the role of past and future. The fact that this wave propagates from future into past is the seemingly counterintuitive part, but its requirement is to transfer potential to actual. This constitutes the \emph{Schr\"odinger picture} for the dilated object-apparatus dynamics.

The propagating waves are null, massless, and they continually pass through the boundary and thus the supposed object. The quantum object and the massless waves form a separable compound state when considered formally as mathematics.
At random instants the object and the waves are considered to spontaneously interact, and each interaction generates a derivative of the object. Perhaps the most interesting thing is that the interaction, a measurement, results in the entanglement of the object and the apparatus in such a way that the object becomes inseparable from the past.

%Further, if one considers the object as a wave-function for a position operator, then one may suppose that the measurement of this object, by the propagating massless flux (a wave over a time-line with two degrees of freedom), results in a wave-function for a position operator which is no longer separable from time parameter of the evolution.

\section{Irreducible Representation of $\mathrm{d}t$}
%\begin{theorem} Let $\varphi$ be a chaotic state over a Newton-Leibniz monoid $\mathfrak{m}$, having the exponential form $\varphi(g)=\exp\{\lambda(g)\}$ for an infinitely divisible state $\lambda(g)=\int l_x(g)\mathrm{d}x$, $l_x(f\star h)^\ast=l_x(h\star f)$, then there exists a Minkowski-Hilbert space $\Bbbk_x=\big\{\mathbb{C}^2,\boldsymbol{\eta}\big\}$, a unital $\star$-representation, $\pi(f\star g)=\pi(f)\pi(g)^\star$, of $\mathfrak{d}$ in $\mathcal{M}_2(\mathcal{B})$, and a gauge vector $\xi\in\Bbbk$, $\|\xi\|=0$, such that $l_x(g)=\xi^\star\pi_x(g)\xi$. \end{theorem}
The irreducible representation of the deterministic differential increment $\mathrm{d}t$ is one aspect of the algebraic realization of calculus. The complete theory also includes the matrix representations of stochastic differential increments, some of which will be touched upon in later chapters. For now it is essential to obtain a clear understanding of the increment of continuous time as an algebraic object.
The first step to achieving this is to become clear about the properties that a deterministic derivative $\mathrm{d}V=K\mathrm{d}t$ must satisfy in order to be regarded as an element of an involutive algebra (algebra equipped with involution).

The standard definition of the derivative $K$ of a function $V$ may be given as
\[
K(t)=\lim_{\delta\searrow0}\frac{V(t+\delta)-V(t)}{\delta}
\]
corresponding to the differential increment
\[
\mathrm{d}V(t)=V(t+\mathrm{d}t)-V(t)=K(t)\mathrm{d}t
\]
which now attempts to define $\mathrm{d}t$ as an algebraic element. The advantage of defining $\mathrm{dt}$ as an algebraic element is that the expression for $K$ becomes exact, rather than the result of a limit. It also turns out that an algebraic understanding of $\mathrm{d}t$ will provide a new insight into dynamics, in terms of both philosophy and physics.

Generally, a deterministic differential increment $\mathrm{d}V$ is an element of the \emph{Newton-Liebniz algebra} $\mathfrak{d}$, which is a one-dimensional subalgebra of a \emph{general quantum It\^o algebra}. The single dimension refers to the fact that the elements of $\mathfrak{d}$ may be given in terms of a single basis element $\mathrm{d}t$; we shall see that the structure of this basis element is non-trivial and cannot be omitted. However, the coefficient $K$ of the derivative $\mathrm{d}V$ may be many, or even infinite, dimensions regarded as the internal degrees of freedom of the Newton-Leibniz algebra.

The algebra $\mathfrak{d}$ is equipped with multiplication and addition, but it is nilpotent which means that the product of any two elements of the algebra is zero, so notice that $\mathfrak{d}$ is not unital - it does not contain an identity element. The algebra $\mathfrak{d}$ also has an involution $\ast:a\mapsto a^\ast\in\mathfrak{d}$ for all $a\in\mathfrak{d}$.
\begin{remark}When $\mathfrak{d}$ is unitized to form the monoid  $\mathfrak{m}=\boldsymbol{1}+\mathfrak{d}$ the induced product is not nilpotent but corresponds to the addition in $\mathfrak{d}$, $(\boldsymbol{1}+a)(\boldsymbol{1}+b)=\boldsymbol{1}+(a+b)$ for all $a$ and $b$ in $\mathfrak{d}$, but bear in mind that things are a little more complicated in the stochastic generalization.
\end{remark}The nilpotenet property of the Newton-Leibniz algebra is given by the basis element $\mathrm{d}t$, so that $\mathrm{d}t\mathrm{d}t=0$. The basis element is also required to be invariant under the action of the involution such that given $a=K\mathrm{d}t$ in $\mathfrak{d}$ we have $a^\ast=K^\dag\mathrm{d}t$ which is also in $\mathfrak{d}$;  with respect to an involution $\dag$ defined in the space of the coefficients $K$.
\subsection{The Derivation of Minkowski Space}
We may summarize by stating that there are  two properties which the deterministic increment must satisfy, these are
\[
\mathrm{d}t^\ast=\mathrm{d}t,\qquad \mathrm{d}t^\ast\mathrm{d}t=0,
\]
and it is from these two conditions that we arrive at an \emph{irreducible representation} of this basis element of the Newton-Leibniz algebra. This representation is given by a map
\[
\pi:\textrm{\textbf{nilpotent algebra}}\;\mapsto\;\textrm{\textbf{matrix sub-algebra}}
\]
and maps $\mathrm{d}t$ into a one-dimensional subalgebra of the $2\times2$ matrix algebra $\mathcal{M}_2$.
The representation is unique and given in  the canonical form as
\begin{equation}
\pi(\mathrm{d}t)=\left[
                   \begin{array}{cc}
                     0 & 1 \\
                     0 & 0 \\
                   \end{array}
                 \right].\label{1}
\end{equation}
The involution in $\mathfrak{d}$ is represented by a pseudo-involution in the matrix representation because it is induced by a pseudo-metric. In fact it is induced by a non-diagonal form of the Minkowski metric which we shall denote by $\boldsymbol{\eta}$.
We denote this pseudo-involution by $\star$, so that $\pi(\mathrm{d}t^\ast)=\pi(\mathrm{d}t)^\star$, and it has the explicit form
\begin{equation}
\pi(\mathrm{d}t)^\star:=\boldsymbol{\eta}\pi(\mathrm{d}t)^\dag\boldsymbol{\eta}\label{2}
\end{equation}
where $\dag$ denotes the standard matrix involution of transposition and complex conjugation, and
\[
\boldsymbol{\eta}=\left[
                    \begin{array}{cc}
                      0 & 1 \\
                      1 & 0 \\
                    \end{array}
                  \right].
\]
We shall henceforth speak of `$\star$-involution' as distinct from the usual  `$\dag$-involution'.

The fact that the $\star$-involution is induced by a Minkowski metric ultimately comes from
a consideration of states on the nilpotent algebra $\mathfrak{d}$. The canonical matrix representation of the general quantum It\^o algebra, also called $\star$-algebra, was constructed by Belavkin \cite{Be92b} in a manner similar to the GNS construction, but he discovered that $\star$-algebra was fundamentally hyperbolic in the the sense that it is  represented on a \emph{pseudo}-Hilbert space \emph{with Minkowski metric}.

The $\star$-involution may be derived from the construction of states on  $\mathfrak{d}$, which are given by pseudo-Hermitian forms, so that
\[
\langle \xi,\mathbf{K}^\star\xi\rangle= \langle\mathbf{K} \xi,\xi\rangle
\]
for the $\star$-adjointable operators $\mathbf{K}=K\otimes\pi(\mathrm{d}t)$. The pseudo-state vectors $\xi$ are null, $\|\xi\|^2=\langle\xi,\xi\rangle=0$ but may be regarded as unital in the sense that
\[
\langle \xi,\pi(\mathrm{d}t)\xi\rangle=1.
\]
\subsection{The Physical Gauge}
It is from  the canonical representation $\pi(\mathrm{d}V)=\mathbf{K}$ of deterministic increments $\mathrm{d}V$ that one may obtain dynamics from the construction of pseudo-state vectors $\xi$. This begins with the establishment of both a physical input and a physical  output gauge vector. The requirements for such  gauge vectors are simply the logical interpretation of a formality regarding the nature of the time increment $\pi(\mathrm{d}t)$.

The output gauge vector is denoted by $\xi_-$ and the input gauge vector by $\xi_+$, they are both null and required to satisfy the  conditions
\[
\pi(\mathrm{d}t)\xi_+=\xi_-,\qquad \pi(\mathrm{d}t)\xi_-=0,\qquad \langle\xi_-,\xi_+\rangle:=\xi_-^\star\xi_+=1,
\]
and although $\xi_+$ is unique only up to the addition of $\mathrm{i}\phi\xi_-$, where $\phi\in\mathbb{R}$ and $\mathrm{i}=\sqrt{-1}$, we define
\[
\xi_+=\left[
  \begin{array}{c}
    0 \\
    1 \\
  \end{array}
\right],\qquad \xi_-=\left[
  \begin{array}{c}
    1 \\
    0 \\
  \end{array}
\right].
\]

The input gauge vector may also be referred to as the future-state of the apparatus, and the output gauge vector the past-state. This is a consequence of understanding that the arrow of time is encoded in the matrix representation of the time-increment, which as an element of a nilpotent algebra has no inverse: $\pi(\mathrm{d}t)^{-1}$ does not exist (although its unitalization $\boldsymbol{1}+\pi(\mathrm{d}t)$ is invertible). Thus from the action of the time increment we may infer a method by which strict order can be imposed. That means that when given a set of points at which the apparatus has reacted (i.e. measurements have been taken) these points may be ordered into a chain by virtue of the identification of any subsequent point as previously being in the future of all previous points.  This ability to order measurement data requires that the data is stored. We shall see that this arises naturally from the entanglement of the measured object with the past.
\subsection{The Lorentz Representation of Minkowski Space}
 One may have noticed that the Minkowski metric $\boldsymbol{\eta}$ appearing here does not have its familiar diagonal form. However, this non-diagonal form is transformed to the diagonal one, familiar to Special Relativity, by the action of  the $\dag$-unitary operator
\[
           \mathbf{U}=\tfrac{1}{\sqrt2}\left[
                 \begin{array}{cc}
                   1 & 1 \\
                   1 & -1 \\
                 \end{array}
               \right]                      \]
so that
\[
\mathbf{U}^\dag\boldsymbol{\eta}\mathbf{U}=\left[
                                   \begin{array}{cc}
                                     1 & 0 \\
                                     0 & -1 \\
                                   \end{array}
                                 \right].
\]
In accordance with this rotation of the basis in the Minkowski space we may observe that the gauge vectors are transformed to
\[
\mathbf{U}^\dag\xi_+=\frac{1}{\sqrt2}\left[
                 \begin{array}{c}
                   1 \\
                   -1 \\
                 \end{array}
               \right],\quad \mathbf{U}^\dag\xi_-=\frac{1}{\sqrt2}\left[
                 \begin{array}{c}
                   1 \\
                   1 \\
                 \end{array}
               \right].
\]
%which are none other than the Gupta-Bleuler
Further, the \emph{real Lorentz transform is diagonalized} when transformed into the basis of the pseudo-apparatus, such that
\begin{equation}
\mathbf{U}\left[
            \begin{array}{cc}
              \cosh\theta & \sinh\theta \\
              \sinh\theta & \cosh\theta \\
            \end{array}
          \right]\mathbf{U}^\dag=\left[
                                   \begin{array}{cc}
                                     \exp\{\theta\} & 0 \\
                                     0 & \exp\{-\theta\} \\
                                   \end{array}
                                 \right]\equiv \boldsymbol{\upsilon}^\star\label{3}
\end{equation}
giving immediate cause for the identification of the space of the pseudo-apparatus as the Lorentz Representation of Minkowski space. For the sake a verbal communication note that the Lorentz transforms $\boldsymbol{\upsilon}$, whichever representation of Minkowski space is used, are defined as the real \emph{pseudo-unitary} linear operators $\boldsymbol{\upsilon}^\star=\boldsymbol{\upsilon}^{-1}$ where the pseudo-involution is induced by the Minkowski metric of that representation.
\section{Evolution of a Quantum System}
So far we have established that a deterministic increment $\mathrm{d}V=K\mathrm{d}t$ may be dilated to a nilpotent operator $\mathbf{K}$. This dilation is not a way of building in `by hands' any concepts about physics, but rather it is a natural consequence of the study of $\mathrm{d}t$ as an algebraic object from which physical concepts follow. This article is intended to introduce the subject matter to a wider audience, but the author refers readers to both \cite{MFB11} and \cite{Be92b} for a more thorough treatment of the matrix representation of $\mathrm{d}t$ and the necessity of Minkowski space in this representation.

Our next task is to consider $V$ as an evolution propagator for a quantum system. To do this we shall consider the case when $V$ is an operator on an arbitrary Hilbert space $\mathfrak{h}$ to which properties of a quantum system are referred. An arbitrary state of the quantum system shall be denoted by $\psi$, a unit vector in $\mathfrak{h}$. The pseudo-Hilbert space of the apparatus shall be denoted by $\Bbbk$, this is generally a complex Minkowski space and shall be referred to as \emph{Minkowski-Hilbert space}.
\subsection{Spontaneous Interaction}
In the Minkowski-Hilbert product space $\mathfrak{h}\otimes\Bbbk$ we may consider a compound state (pseudo-state in fact) of the object plus apparatus having the separable form $\psi\otimes\xi_+$; the apparatus is prepared and the quantum object is simply considered. A measurement of the quantum object is represented by the action of an operator $\mathbf{G}$ on the state $\psi\otimes\xi_+$. In order for $\mathbf{G}$ to sensibly describe an interaction between the object and apparatus, that ultimately generates evolution in $\mathfrak{h}$, it must have a very particular form. Such a form may be derived unambiguously.

A differential change in an evolution propagator $V$ is indeed denoted by $\mathrm{d}V$, but the incrementally evolved propagator is of course $V+\mathrm{d}V$. This mathematical object is not an element of the algebra $\mathfrak{d}$ but instead has the canonical dilation
\[
V+\mathrm{d}V\mapsto V\otimes \mathbf{I}+\mathbf{K}
\]
where $\mathbf{I}$ is the  $2\times2$ identity matrix and $\mathbf{K}=\pi(\mathrm{d}V)\equiv K\otimes\pi(\mathrm{d}t)$. In particular, in the case when the propagator $V$ has exponential form corresponding to a generator $L$, so that $K=LV$, we find that $V+\mathrm{d}V=(\boldsymbol{1}+L\mathrm{d}t)V$ which has the canonical dilation
\begin{equation}
V+\mathrm{d}V\mapsto\left[
  \begin{array}{cc}
    I & L \\
    0 & I \\
  \end{array}
\right]\odot V:=\left[
  \begin{array}{cc}
    V & LV \\
    0 & V \\
  \end{array}
\right]\label{4}
\end{equation}
where we have used the symbol $\odot$ for the semi-tensor product which behaves as tensor product for the apparatus and non-commutative matrix product for the quantum object.

It should now be apparent that the interaction operator $\mathbf{G}$ that describes the process of measurement of the quantum object, by the pseudo-apparatus, is given as
\begin{equation}
\mathbf{G}=\left[
  \begin{array}{cc}
    I & L \\
    0 & I \\
  \end{array}
\right].\label{5}
\end{equation}
To see this one may first set $V=I$, which follows from  the arbitrariness of the state $\psi$. Then we find that the new state of the object resulting from a differential increment is
\[\psi+\mathrm{d}\psi=(\boldsymbol{1}+L\mathrm{d}t)\psi.
\]
In the context of the pseudo-apparatus we can also see that when this \emph{differential transition} is dilated (i.e. represented as a matrix) the operator $\mathbf{G}$, defined in (\ref{5}), is precisely that which generates this transition. So we shall now study the role of such an operator $\mathbf{G}$ as a transformation of the compound state $\psi\otimes\xi_+$.

First of all, notice that the interaction $\mathbf{G}(\psi\otimes\xi_+)$ has the explicit form
\[
\left[
  \begin{array}{cc}
    I & L \\
    0 & I \\
  \end{array}
\right]\left[
         \begin{array}{c}
           0 \\
           \psi \\
         \end{array}
       \right]=\left[
         \begin{array}{c}
           L\psi \\
           \psi \\
         \end{array}
       \right]
\]
but, in particular, we have a dilation of the \emph{differential increment}
of the quantum object given as
\[
\mathbf{L}:\left[
                           \begin{array}{c}
                             0 \\
                             \psi \\
                           \end{array}
                         \right]\mapsto
\left[
                           \begin{array}{c}
                             L\psi \\
                             0 \\
                           \end{array}
                         \right]\qquad\textrm{where}\qquad \mathbf{L}:=\left[
  \begin{array}{cc}
    0 & L\\
    0 & 0 \\
  \end{array}
\right]
\]
so that we may establish a more detailed mathematical construction of the idea of change in a system. That is, the system is said to \emph{change} if
\[
\psi\mapsto L\psi\qquad\textrm{and}\qquad \left[
                           \begin{array}{c}
                             0 \\
                             1 \\
                           \end{array}
                         \right]\mapsto\left[
                           \begin{array}{c}
                             1 \\
                             0 \\
                           \end{array}
                         \right].
\]
Thus the transformation $L\psi$ of $\psi$ is inferred as the result of the pseudo-measurement outcome $\xi_+\mapsto\xi_-$. The operator $L$ is generally an integrable operator densely defined over some time-like domain as a bounded operator in the Hilbert space $\mathfrak{h}$. It is enough to consider it here as a constant, bounded operator on a bounded interval of such a time-like domain.

The final thing to notice here is that although the differential operator $\mathbf{L}$ does not define an entangling operation, the interaction operator $\mathbf{G}=\mathbf{I}+\mathbf{L}$ does. Formally, this entanglement requires two additional things. The first is that the measurement process is `second-quantized', which corresponds to a sequential object-apparatus interaction dynamics, and the second is that the operator $\mathbf{G}$ be made $\star$-unitary. The latter gives rise to the  generator $L=-\mathrm{i}H$ given by a self-adjoint Hamiltonian $H$.
\subsection{Sequential Interaction Dynamics}
The action of $\mathbf{G}$ on the vector $\psi\otimes\xi_+$ corresponds to a single object-apparatus interaction. For a sequence of such object-apparatus interactions we may write down a stochastic differential equation that is given with respect to the stochastic differential $\mathrm{d}n$ of a counting process. It is not our intention to go into to many details about such a counting process. Here it is enough to understand that $\mathrm{d}n$ takes only the values 0 or 1. In fact it is zero unless there is a spontaneous interaction between the object and the apparatus, in which case $\mathrm{d}n=1$.

The idea of spontaneous interaction may be a topic of debate in physics, but here it is understood in terms of its derivation from a differential increment. So it is not necessary to justify the existence of a dynamical process that corresponds to the counting  of a sequence of spontaneous interactions. However, one may refer to \cite{Be00b} for Belavkin's explanation of the  existence of spontaneous `jumps' in a system, resulting from measurement, as `the Bayesian selection rule of a posterior state from a prior mixture of states corresponding to possible measurement results'. Bear in mind that in the context of this pseudo-measurement we have a selection rule for the derivative of a state (which is a pseudo-state) from a trivial mixture of this derivative with zero - a representation of the outcome of no measurement and thus no derivative of the state of the quantum system.

The dynamical equation for a sequence of pseudo-measurements of a quantum object is formally given on finite chains \[
\vartheta=\{t_1<t_2<\ldots <t_n\}.
 \]The points $t_i$ simply attribute a coordinate to an interaction and the domain of such coordinates is a continuous domain in which pairs of coordinates may be strictly ordered. If the domain is not assumed to be continuous then we encounter a contradiction, because this continuity follows from the assumption that we are considering a differentiable function $V$ (the evolution propagator of the quantum object). So, as a canonical dilation of deterministic dynamics, which assumes differentiability of $V$, we may quite rightly consider a sequence of interactions at a set of points in $\mathbb{R}$. In fact, without loss of generality we can take $t_i\in\mathbb{R}_+$.
%\begin{remark}Indeed it is a topic of interest to the author to consider Hamiltonian dynamics as something purely derived from the discrete interaction dynamics. It is not a problem to establish an ordering relation on interactions based on a fundamental principle of past and future, but justification of a continuous domain of such orderable interactions is not so straight forward.  \end{remark}

We shall now denote by $\psi_0:=\boldsymbol{F}^\star\psi$ the initial state of the combined system of the quantum object plus second quantized apparatus. Indeed $\psi$ represents the initial state of the object, whilst $\boldsymbol{F}^\star$ is an embedding operator which composes the object with the initial state of the second-quantized apparatus. The second-quantization of the apparatus is simply a consideration of any finite number of copies of the apparatus, including zero, and $\boldsymbol{F}^\star$ has each copy of the apparatus prepared in the \emph{future/input} state.

Formally the embedding operator is defined for any finite chain $\vartheta$ as a map
\[
\boldsymbol{F}^\star(\vartheta):\psi\mapsto\psi\otimes\xi_+(t_1)\otimes\xi_+(t_2)\otimes\cdots\otimes\xi_+(t_n)
\]
which is given with respect to $n$ copies of the input gauge vector which we now index by the coordinates $t_i\in\vartheta$. One should note that $\boldsymbol{F}^\star$ is isometric, $\boldsymbol{FF}^\star=I$ is the identity on $\mathfrak{h}$, and this is also a statement that the second quantization of the apparatus null-vector is no longer null but normalized. We shall also make the notational identification $\psi=\psi(0)$ so that the initial compound state vector has the evaluations $\psi_0(\vartheta)=\boldsymbol{F}^\star(\vartheta)\psi(0)$.

In terms of the chain $\vartheta$ we may also give more explicit definition to the counting differential so that
\[
\mathrm{d}n_t(\vartheta)=1\quad\textrm{if}\quad t\in\vartheta,\quad\textrm{otherwise}\quad \mathrm{d}n_t(\vartheta)=0\quad\textrm{if}\quad t\notin\vartheta
\]
and now we have an understanding of all the necessary components we may write down the equation of counting dynamics, counting sequential measurements of the quantum object by the pseudo-apparatus. This is
\begin{equation}
\mathrm{d}\psi_t(\vartheta)=\mathbf{L}\psi_t(\vartheta)\mathrm{d}n_t(\vartheta),\qquad\psi_0(\vartheta)= \boldsymbol{F}^\star(\vartheta)\psi(0),\label{6}
\end{equation}
and this equation reads as: \emph{a change in the state of the object-apparatus system is given by the action of the differential operator $\mathbf{L}$ at arbitrary points $t\in\vartheta$.}
The solution $\psi_t(\vartheta)$ describes an entanglement of the object with the apparatus on $[0,t)\subset\mathbb{R}_+$ and this entanglement is a transformation given by a %\emph{semi-tensor}
second-quantization of the interaction operators $\mathbf{G}=\mathbf{I}+\mathbf{L}$, denoted by $\mathbf{G}^\odot$, so that
\[
\psi_t(\vartheta)=\mathbf{G}_t^\odot(\vartheta)\boldsymbol{F}^\star(\vartheta)\psi(0)
\]
where $\mathbf{G}_t^\odot(\vartheta)=\odot_{z\in\vartheta^t}\mathbf{G}(z)$, $\vartheta^t=\vartheta\cap[0,t)$. Here $\mathbf{G}(z)=\mathbf{G}$ as we are assuming that the generator is constant, but in a more general setting \cite{MFB11} one may consider a different interaction operator at each point in a given chain $\vartheta$.
\begin{remark}
One may wish to note explicitly that
\[
\mathbf{G}\odot\mathbf{G}=\mathbf{I}-H^2\otimes\pi(\mathrm{d}t)\otimes\pi(\mathrm{d}t)
\]
where we have simply used $\mathbf{I}$ to denote the identity on $\mathfrak{h}\otimes\Bbbk\otimes\Bbbk$. If the semi-tensor product $\odot$ were replaced with the tensor product $\otimes$ then we would have $H\otimes H$ in place of $H^2$. On the other hand, if $\mathfrak{h}\cong\mathbb{C}$ then $\odot$ and $\otimes$ would coincide.
\end{remark}
\section{Projecting Out a Deterministic Dynamics}
The second quantization that we are using is given with respect to a Hilbert space called Guichardet-Fock space \cite{Gui72}. We shall not go into many of the details of this here, but there are some things worthy of note. The first is a reiteration that although the apparatus vector $\xi$ is null, its second-quantization $\xi^\otimes$ is not. This is because second quantization is basically an exponentiation of that which is being second quantized, but there are more general vectors in the second-quantization $\mathbb{F}=\Gamma(\Bbbk)$ of $\Bbbk$ that are still null.
In addition to this we shall establish some basic properties of the pseudo Guichardet-Fock space $\mathbb{F}$, which shall be called Minkowski-Fock space.

\emph{Product vectors} in $\mathbb{F}$ are denoted by $\xi^\otimes$ for some $\xi\in\Bbbk$, and they may be normalized to form \emph{coherent} vectors. These product vectors are also called \emph{exponential vectors} because their norm has the exponential form
\[
\|\xi^\otimes\|^2=\exp\{\xi^\star\xi\}.
\]
At this stage we must be a little more careful about our understanding of these spaces. The Minkowski-Hilbert space $\Bbbk$ is actually the space $L^1(\mathbb{R}_+)\oplus L^\infty(\mathbb{R}_+)$ and the adjoint space $\Bbbk^\star$ is $L^\infty(\mathbb{R}_+)\oplus L^1(\mathbb{R}_+)$, but details of this derivation are beyond the scope of this article. So at any point $t\in\mathbb{R}_+$ we have, as introduced earlier for example
\[
\xi^\star_+(t)\pi(\mathrm{d}t)\xi_+(t)=1
\]
but now we must understand the vectors $\xi\in\Bbbk$ more specifically as vector functions $\xi:t\mapsto\xi(t)$ in $L^1(\mathbb{R}_+)\oplus L^\infty(\mathbb{R}_+)$ so that
\[
\xi^\star\xi=\int_0^\infty\xi^\star(t)\xi(t)\mathrm{d}t.
\]
Indeed there may appear to be a contradiction in the sense that we are using calculus to understand calculus. However, what we are actually doing is deriving the precise mechanism that represents differential increments and showing that the \emph{expectation} of this mechanism produces integration. It was stated earlier that this is to provide insight into physics and philosophy. As far as the mathematics is concerned the integration in the pseudo inner-product $\xi^\star\xi$ is simply defined in terms of the Lebesgue measure on $\mathbb{R}_+$.
\subsection{An Expectation of The Measurement Process}
The evolution propagator $V$ resolving the Schr\"odinger equation
\begin{equation}
\mathrm{d}V(t)=-\mathrm{i}HV(t)\mathrm{d}t,\qquad V(0)=I,\label{8}
\end{equation}
may be obtained as a projection of the sequential interaction operation $\mathbf{G}^\odot_t$ into the algebra of operators on the quantum-object Hilbert space $\mathfrak{h}$. This is an expectation
\begin{equation}
V(t)=\verb"E"\big[\mathbf{G}^\odot_t\big]\equiv\boldsymbol{F}\mathbf{G}^\odot_t\boldsymbol{F}^\star\label{9}
\end{equation}
for the case when the object-apparatus interaction has the form
\begin{equation}
\mathbf{G}=\left[
             \begin{array}{cc}
               I & -\mathrm{i}H \\
               0 & I \\
             \end{array}
           \right],\label{10}
\end{equation}
where $H$ is the Hamiltonian that is densely defined as a bounded self-adjoint operator on $\mathfrak{h}$. The expectation is obtained by tracing over the second-quantized apparatus Hilbert space $\mathbb{F}$. This expectation amounts to an averaging over all numbers of interactions (points in $\vartheta$), and for any fixed number of interactions it averages over all possible configurations of the interactions preserving  the order of the points in $\vartheta$. In this way it may be understood as a Feynmann path integral, but perhaps more appropriately it may
simply be understood as the dynamics that appears `in the ignorance of the measurement process' \cite{Be00b}.
\begin{remark}
{It is worth noting that Lindblad dynamics is simply a more general example of this, in which the measurement process has additional degrees of freedom - which are the indices summed over in the Lindblad equation. These additional degrees of freedom correspond the standard quantum noises.}
\end{remark}

Notice that the  interaction operator (\ref{10}) is $\star$-unitary, $\mathbf{G}^\star\mathbf{G}=\mathbf{I}=\mathbf{G}\mathbf{G}^\star$, recalling that $\mathbf{G}^\star=\boldsymbol{\eta}\mathbf{G}^\dag\boldsymbol{\eta}$. In fact the evolution propagator $V$ is a $\dag$-unitary operator if and only if the interaction operator $\mathbf{G}$ is $\star$-unitary. This statement also holds in the case when $V$ is more generally a quantum  stochastic process. In such cases $\mathbf{G}$ is, externally, a $3\times3$ upper-triangular matrix; see \cite{Be92,Be92b}.
\subsection{The Poisson Process and Lorentz Transformations}
Recall that the Lorentz transform $\boldsymbol{\upsilon}$ is also a $\star$-unitary operator, but it does not satisfy the requirements of an interaction operator in the context of this pseudo-measurement dynamics as the interaction operator was required to coincide with the identity matrix on the diagonal. %However, one may re-define the notion of a unit when unitalizing the nilpotent algebra $\mathfrak{d}$ to obtain $\mathfrak{m}_\lambda=\boldsymbol{\upsilon}_\lambda+\mathfrak{d}$ but this does not provide
However, the Lorentz transforms do define an automorphism of the monoid $\mathfrak{m}=\boldsymbol{1}+\mathfrak{d}$ (which is where the interaction operators live) leaving the unit invariant and `boosting' the representation $\pi(\mathrm{dt})$ of the deterministic increment to $\pi_\nu(\mathrm{d}t):=\nu\pi(\mathrm{d}t)$, where $\nu$ is a positive real number.

To understand this one should recall that the Lorentz transform is a diagonal operator that may assume the form
\[
\boldsymbol{\upsilon}_\nu=\left[
                            \begin{array}{cc}
                              \frac{1}{\sqrt{\nu}} & 0 \\
                              0 & \sqrt{\nu} \\
                            \end{array}
                          \right]
\]
from which it may be seen that $\boldsymbol{\upsilon}_\nu^\star\boldsymbol{\upsilon}_\nu=\mathbf{I}$ and $\boldsymbol{\upsilon}_\nu^\star\pi(\mathrm{d}t)\boldsymbol{\upsilon}_\nu=\nu\pi(\mathrm{d}t)$. Notice that the Lorentz transform % should be regarded as both an operator on the object and the apparatus although it appears to be
is primarily operating in the apparatus space which is where the object's time comes from. If one wishes to understand this as some kind of relativistic transformation of the apparatus then one should also bear in mind the effect that it has on the Schr\"odinger equation. For we find that the expectation of the Lorentz transformed dynamics is the solution of a \emph{boosted Schr\"odinger equation}
\begin{equation}
\mathrm{d}V_\nu(t)=-\mathrm{i}HV_\nu(t)\nu\mathrm{d}t,\qquad V_\nu(0)=I.\label{11}
\end{equation}
The propagator $V_\nu$ is obtained as the expectation of the Lorentz transformed interaction operator $\boldsymbol{\upsilon}^\star_\nu\mathbf{G}\boldsymbol{\upsilon}_\nu=\mathbf{I}-\mathrm{i}H\pi_\nu(\mathrm{d}t)$ as
\[
V_\nu(t)=\verb"E"\big[(\boldsymbol{\upsilon}^\star_\nu\mathbf{G}_t\boldsymbol{\upsilon}_\nu)^\odot\big]:= \boldsymbol{F}_\nu\mathbf{G}_t^\odot\boldsymbol{F}^\star_\nu
\]
where $\boldsymbol{F}_\nu$ is simply a second-quantization of the Lorentz transform of the apparatus input state, $\boldsymbol{\upsilon}_\nu\xi_+$.

We would like to understand the meaning of this Lorentz transform in the context of a stochastic process. In fact there is a very nice connection. This boosted dynamics is actually a Poisson expectation of the dilated interaction dynamics (\ref{6}). That is
\begin{equation}
V_\nu(t)=\verb"P"^t_{\nu}\big[\mathbf{G}_t^\odot\big],
\end{equation}
 where $\verb"P"^t_{\nu}\big[\;\cdot\;]$ is the Poisson expectation defined by the Poisson measure which is given with respect to the chains $\vartheta\subset[0,t)$ as
\[
\mathrm{d}\verb"P"^t_{\nu}(\vartheta)=(2\nu)^\otimes(\vartheta)\exp\{-2\nu t\}\mathrm{d}\vartheta,
\]
and by the Poisson state-vectors $\frac{1}{\sqrt2}(1,1)$ in $L^1[0,t)\oplus L^\infty[0,t)$. See \cite{MFB11} for more details.

This means that the discrete interaction dynamics (\ref{6}) of the pseudo-measurement process may be considered as a Poisson process with intensity $\nu$, which is simply the intensity of the object-apparatus interaction process. The factor 2 appears non-trivially due to the fact that although the algebra $\mathfrak{d}$ is externally one dimensional it is irreducibly represented in  two dimensions. Although the reader will notice that this factor does not appear in the Boosted Schr\"odinger equation (\ref{11}).
\subsection{Stochastic Input, Deterministic Output}
The pseudo-Poisson process is described by the stochastic differential equation (\ref{6}) but with initial condition $\psi_0=\boldsymbol{F}_\nu^\star\psi(0)$. This is describing the object-apparatus interaction dynamics which hopefully, by now, the reader is becoming familiar with. Only now we have a notion of the frequency at which these interactions occur - this is the intensity of this Poisson process. In particular, (\ref{6}) may be understood as a Poisson process with unit intensity.

Our final task in obtaining our desired picture  (which is called the Schr\"odinger picture for this stochastic process \cite{Be00a}) is to define a unitary shift of the apparatus. The apparatus may generally be understood from its wave-function $\xi$. The picture we have at present is one in which the quantum object spontaneously interacts with this wave-function at different points $t$ in a random chain $\vartheta$ (a stochastic trajectory of the apparatus). What we would like to do now is consider the apparatus wave-function to be propagating towards the quantum object. This means that we must extend the domain of $\vartheta$, and thus the Minkowski-Hilbert space, onto $\mathbb{R}$. This allows us to define a $\star$-unitary shift which describes a propagation of the apparatus wave-function over $\mathbb{R}$, from the positive into the negative.

This picture has the quantum object fixed at the origin of $\mathbb{R}$ in the sense that all interactions of the object and the apparatus occur at the point $0\in\mathbb{R}$. At this stage, the domain $\mathbb{R}$ may be identified with a coordinate-time for the object in which a set of potential interaction-coordinates, $\vartheta$, are initially considered to be in the future of the object. As the incoming measurement waves propagate towards the object the random interaction coordinates may be considered to propagate as $\vartheta-t=\{t_1-t<t_2-t<\cdots\}$ where $t$ is the evolution parameter of the object. Indeed, whenever $t=t_i\in\vartheta$ an interaction is considered, and this coincides with $t_i-t=0$. These waves come from the object's future.

The shift operator is defined as $\mathbf{T}^t=\exp\{t\partial_\vartheta\}$, where $\partial_\vartheta =\sum_{z\in\vartheta}\partial_z$, and it does not alter the propagator resolving the Schr\"odinger equation,
\[
\verb"P"^t_{\nu}\big[\mathbf{G}^\odot\big]= \verb"P"^t_{\nu}\big[{\mathbf{T}^t}^\star\mathbf{G}^\odot\mathbf{T}^t\big],
\]
so we do not have to worry about anything appearing in the Schr\"odinger equation due to the choice of this picture. However, this picture is very interesting because it is described mathematically by the Schr\"odinger boundary value problem
\begin{equation}
\partial_t\psi^t_\nu(\vartheta)=\partial_\vartheta\psi^t_\nu(\vartheta),\qquad \psi^t_\nu(0_-)=\mathbf{G}\psi^t_\nu(0_+),
\end{equation}
when $\nu$ is constant. The indices $\mp$ are used for the boundary coordinate $0\in\vartheta$ to indicate that the function $\psi^t_\nu(\vartheta)$ has a discontinuity at $0\in\vartheta$. This means that there is a discontinuity in the values of $\psi^t_\nu$ at the boundary between the past (-) and the future (+), and these two values are connected by the operator $\mathbf{G}$.

 First of all, from this Schr\"odinger boundary value problem comes the structure of a second-quantized Dirac boundary value problem. This appears when the dynamics is transformed to a reflection at the boundary of $\mathbb{R}_+$ \cite{MFB11,Be00a}, so that
\begin{equation}
\left[
                                            \begin{array}{cc}
                                              \partial_t+\partial_\vartheta & 0 \\
                                              0 & \partial_t-\partial_\vartheta \\
                                            \end{array}
                                          \right]\left[
                                                   \begin{array}{c}
                                                     \psi^t_-(\vartheta) \\
                                                     \psi^t_+(\vartheta) \\
                                                   \end{array}
                                                 \right]=0,\qquad \psi^t_-(0)=\mathbf{G}\psi^t_+(0),
\end{equation}
where $\psi_+^t$ is the restriction of $\psi^t_\nu$ to the positive domain $\vartheta\subset\mathbb{R}_+$, and $\psi_-^t$ is the reflection of the restriction of $\psi^t_\nu$ to the negative domain $\vartheta\subset\mathbb{R}_-$.

Secondly,in light of the Schr\"odinger boundary value problem we may consider a conditioning of the input Poisson process, which is stochastic, by the $\star$-orthoprojector
\[
\mathbf{P}^t_\nu={\mathbf{T}^t}^\star\boldsymbol{F}_\nu^\star\boldsymbol{F}_\nu^{}\mathbf{T}^t
\]
giving rise to the canonical embedding of a deterministic output flow in $\mathfrak{h}$ such that
\begin{equation}
\mathbf{P}^t_\nu\psi^t_\nu\equiv\mathbf{P}^t_\nu\mathbf{G}^\otimes_t\mathbf{T}^t\boldsymbol{F}_\nu^\star\psi =\mathbf{T}^t\boldsymbol{F}_\nu^\star\psi_\nu(t),
\end{equation}
where $\psi_\nu(t)=V_\nu(t)\psi$ resolves the boosted Schr\"odinger equation.
\section{Large Number Limit of The Pseudo-Measurement Process}
As well as obtaining the Schr\"odinger equation  (\ref{8}) as the conditional expectation of the pseudo-measurement process (\ref{6}) we may also obtain it as the large number limit $\nu\nearrow\infty$ of the measurement process  with initial condition $\psi_0(\vartheta)= \boldsymbol{F}_\nu^\star(\vartheta)\psi(0)$ and transformed generator $\boldsymbol{\upsilon}_\nu\mathbf{G}\boldsymbol{\upsilon}_\nu^\star=\mathbf{G}^\nu$ corresponding to the Dirac boundary value problem \begin{equation}\left[
                                            \begin{array}{cc}
                                              \partial_t+\partial_\vartheta & 0 \\
                                              0 & \partial_t-\partial_\vartheta \\
                                            \end{array}
                                          \right]\left[
                                                   \begin{array}{c}
                                                     \psi^t_-(\vartheta) \\
                                                     \psi^t_+(\vartheta) \\
                                                   \end{array}
                                                 \right]=0,\qquad \psi^t_-(0)= \mathbf{G}^\nu \psi^t_+(0).
                                                 \end{equation}
To see this we may write the dynamics of the measurement process equivalently in the form (\ref{6}) as
\begin{equation}
\mathrm{d}\psi^\nu_t(\vartheta)=\nu^{-1} \mathbf{L}\psi^\nu_t(\vartheta)\mathrm{d}n_t(\vartheta),\qquad\psi^\nu_0(\vartheta)= \boldsymbol{F}^\star_\nu(\vartheta)\psi(0),\label{60}
\end{equation}
where $\mathbf{L}$ is of course $-\mathrm{i}H\otimes\pi(\mathrm{d}t)$. Next we shall re-write (\ref{60}) in the form
\begin{equation}
\mathrm{d}\psi^\nu_t(\vartheta)=-\frac{\mathrm{i}}{\nu}{H} \psi^\nu_t(\vartheta)\mathrm{d}\mathbf{n}_t(\vartheta) \label{61}
\end{equation}
where $\mathrm{d}\mathbf{n}_t:=\pi(\mathrm{dt})\mathrm{d}n_t$ may formally be called a pseudo-counting differential increment; it is nilpotent due to the nilpotency of $\pi(\mathrm{d}t)$.

Now we shall transform to the Fock-vacuum basis which may be achieved by a $\star$-unitary pseudo-Weyl operator $\mathbf{W}_\nu$ so that the initial condition  $\psi^\nu_0(\vartheta)= \boldsymbol{F}^\star_\nu(\vartheta)\psi(0)$ may be re-written as $\psi^\nu_0(\vartheta)= [\mathbf{W}_\nu\delta_\emptyset](\vartheta)\psi(0)$ where
\begin{equation}
\delta_\emptyset(\vartheta)=0\;\textrm{if}\;\vartheta\neq\emptyset,\qquad \delta_\emptyset(\emptyset)=1,
\end{equation}
and the solution $\big(\boldsymbol{\upsilon}_\nu\mathbf{G}_t\boldsymbol{\upsilon}^\star_\nu\big)^\odot$ of (\ref{61}) is transformed to $\mathbf{W}^\star_\nu\big(\boldsymbol{\upsilon}_\nu\mathbf{G}_t\boldsymbol{\upsilon}^\star_\nu\big)^\odot \mathbf{W}_\nu$. Ultimately this allows us to  write (\ref{61}) as
 \begin{equation}
\mathrm{d}\check{\psi}^\nu_t(\vartheta)+{\mathrm{i}}{H} \check{\psi}^\nu_t(\vartheta)\mathrm{d}t= -\frac{\mathrm{i}}{\nu}{H} \check{\psi}^\nu_t(\vartheta)\mathrm{d}\widetilde{\boldsymbol{m}}^\nu_t(\vartheta) \label{62}
\end{equation}
where $\check{\psi}^\nu_t=\mathbf{W}_\nu\psi^\nu_t$ and in particular we find that in the large number limit
\[
\lim_{\nu\nearrow\infty}\;\frac{1}{\nu}\mathrm{d}\widetilde{\boldsymbol{m}}^\nu_t=0
\]
and we recover the Schr\"odinger equation
\[
\mathrm{d}\psi(t)=-{\mathrm{i}}{H} \psi(t)\mathrm{d}t.
\]
As the wave-function $\check{\psi}^\infty_t(\vartheta)$ is not depending on $\vartheta$ it is unique up to the choice of initial condition. So we may simply define $\check{\psi}_0^\infty=\boldsymbol{F}^\star\psi(0)$.

%deduce that the pseudo-counting increment $\mathrm{d}\mathbf{n}_t$ has the expectation $\nu\mathrm{d}t$, but it does require that we dilate the stochastic differential equation (\ref{62}) first.

%the transformation $\mathbf{G}^\nu\boldsymbol{\upsilon}_\nu\xi_+$ of the input gauge vector $\boldsymbol{\upsilon}_\nu\xi_+$ such that we find, indeed, that
%\begin{equation}
%\langle\xi_+^\star\boldsymbol{\upsilon}_\nu^\star{\mathbf{G}^\nu}^\star \mathrm{d}\mathbf{n}_t \mathbf{G}^\nu\boldsymbol{\upsilon}_\nu\xi_+\rangle=\nu\langle\mathrm{d}n\rangle=\nu\mathrm{d}t
%\end{equation}

%In order to establish the large number limit of the dynamical process described by (\ref{61}) the dynamics may be dilated a second time. This will allow us to apply the large number limit techniques introduced by Belavkin, which may be seen in  \cite{Be93,Be00b}.
\subsection{The Matrix Representation of $\mathrm{d}n$}
Previously, in (\ref{6}), we did not need to represent the counting increment $\mathrm{d}n$ as a matrix, but now we do so that we can illustrate a method by which the large number limit can be obtained \cite{Be93}. When dilating $\mathrm{d}n$ we must use a $3\times3$ matrix algebra \cite{Be92,Be92b,Be00b}, because the multiplication laws of $\mathrm{d}n$ and $\mathrm{d}t$ must satisfy
\[
\mathrm{d}n\mathrm{d}n=\mathrm{d}n,\quad \mathrm{d}t\mathrm{d}t=0, \quad \mathrm{d}n\mathrm{d}t=0 =\mathrm{d}t\mathrm{d}n.
\]
One may obtain the irreducible matrix representation of $\mathrm{d}n$ as
\[
\varpi(\mathrm{d}n)=\left[
  \begin{array}{ccc}
    0 & 0 & 0 \\
    0 & 1 & 0 \\
    0 & 0 & 0 \\
  \end{array}
\right]\]
and thus the pseudo-counting increment $\mathrm{d}\mathbf{n}$ may be represented as $\pi(\mathrm{d}t)\otimes\varpi(\mathrm{d}n)$. Note that in the $3\times3$ matrix algebra $\mathrm{d}t$ has the form \[\varpi(\mathrm{d}t)=\left[
  \begin{array}{ccc}
    0 & 0 & 1 \\
    0 & 0 & 0 \\
    0 & 0 & 0 \\
  \end{array}
\right].
\]
The reason why we have denoted this $\star$-algebra representation by $\varpi$ is so not to confuse it with the representation $\pi$ of the nilpotent algebra introduced in previous chapters, which  only needed to be represented in a $2\times2$ matrix algebra.

 The pseudo-Weyl operator $\mathbf{W}_\nu$ is obtained as the conditional expectation of a $\star$-unitary operator $\mathscr{Z}_\nu$ as
 \begin{equation}
 \mathbf{W}_\nu=\boldsymbol{F}\mathscr{Z}_\nu^\otimes\boldsymbol{F}^\star
 \end{equation}
 where
 \begin{equation}
 \mathscr{Z}_\nu=\left[
               \begin{array}{ccc}
                 I & -\sqrt{\nu}\xi_+^\star & 0 \\
                 0 & \mathbf{I} &\sqrt{\nu}\xi_+ \\
                 0 & 0 & I \\
               \end{array}
             \right] \label{65}
 \end{equation}
 which transforms the pseudo-counting increment $\varpi(\mathrm{d}\mathbf{n}_t)$ to
 \[
 \mathscr{Z}_\nu^\star\varpi(\mathrm{d}\mathbf{n}_t)\mathscr{Z}_\nu=\left[
               \begin{array}{ccc}
                 0 & \sqrt{\nu}\xi_-^\star & \nu \\
                 0 & \pi(\mathrm{d}t) &\sqrt{\nu}\xi_- \\
                 0 & 0 & 0 \\
               \end{array}
             \right]:=\varpi(\mathrm{d}\widetilde{\boldsymbol{m}}_t^\nu)+\nu\varpi(\mathrm{d}t)
 \]
 which is still a counting increment due to the $\star$-unitarity of $\mathscr{Z}_\nu$.

\section{A Pseudo-Diffusion Equation}
Now that we have established the form of the dilation of Schr\"odinger dynamics as a pseudo-measurement process we can investigate such mathematical techniques as \emph{partial dilation}. To do this we shall first
 consider a Schr\"odinger equation having the form
\begin{equation}
\mathrm{d}\psi(t)=-\mathrm{i}\Big(\lambda-\frac{\nu}{\lambda}\Big)H\psi(t)\mathrm{d}t\label{20}
\end{equation}
for real parameters $\lambda,\nu>0$.
This admits a decomposition of the generator into two parts, and what we can do now is dilate the
dynamics described by (\ref{20}) with respect to just one of these parts. In this way we are able to obtain the stochastic differential equation
\begin{equation}
\mathrm{d}\psi^\nu_t(\vartheta)+{\mathrm{i}}H\psi^\nu_t(\vartheta){\lambda}\mathrm{d}t =\frac{\mathrm{i}}{\lambda}\mathbf{H} \psi^\nu_t(\vartheta)\mathrm{d}n_t(\vartheta),\qquad \psi_0^\nu=\boldsymbol{F}_\nu^\star\psi\label{21}
\end{equation}
where $\mathbf{H}=H\otimes\pi(\mathrm{d}t)$. The first thing to note is that the solution of this equation has the form
\[
\psi^\nu_t(\vartheta)= \exp\{-\mathrm{i} H\lambda t\}(\mathbf{G}^\lambda_t)^\odot(\vartheta)\boldsymbol{F}_\nu^\star(\vartheta)\psi
\]
since $H$ commutes with $\mathbf{G}^\lambda=\mathbf{I}+\frac{\mathrm{i}}{\lambda}\mathbf{H}$. This partially dilated dynamics does indeed describe a pseudo-measurement  process of a quantum system, only this time we have allowed there to by a free evolution of the system in between measurements. This free evolution satisfies the boosted Schr\"odinger equation
\begin{equation}
\mathrm{d}\psi_{free}(t)=-\mathrm{i}H\psi_{free}(t)\lambda\mathrm{d}t.\label{25}
\end{equation}
However, notice that this free evolution also commutes with the projection $\boldsymbol{F}_\nu$ of $\psi^\nu_t$ onto $\mathfrak{h}$, recovering (\ref{20}).

The reason why we have done this is so that we can obtain a well defined central limit of the dynamics described by (\ref{21}).
To do this we shall first re-write (\ref{21}) in the form
\begin{equation}
\mathrm{d}\psi^\nu_t(\vartheta)+{\mathrm{i}}H\psi^\nu_t(\vartheta){\lambda}\mathrm{d}t =\frac{\mathrm{i}}{\lambda}{H} \psi^\nu_t(\vartheta)\mathrm{d}\mathbf{n}_t(\vartheta)\label{22}
\end{equation}
where $\mathrm{d}\mathbf{n}_t=\pi(\mathrm{dt})\mathrm{d}n_t$ as it was in the large number limit above. In order to establish the central limit of the dynamical process described by (\ref{22}) it shall be convenient to represent the increments in (\ref{22}) in their matrix form in a manner similar to the previous section. This will allow us to apply the central limit techniques introduced by Belavkin, which may be seen in  \cite{Be93,Be00b}.

%Here, the deterministic increment appearing in (\ref{22}) must now be considered in a $3\times3$ matrix algebra, because we also have a stochastic increment $\mathrm{d}n$ to represent which will require an extra degree of freedom \cite{Be92,Be92b,Be00b}. The multiplication laws of $\mathrm{d}n$ and $\mathrm{d}t$ are \[ \mathrm{d}n\mathrm{d}n=\mathrm{d}n,\quad \mathrm{d}t\mathrm{d}t=0, \quad \mathrm{d}n\mathrm{d}t=0 =\mathrm{d}t\mathrm{d}n. \] One may obtain the irreducible represents of this two-dimensional $\star$-algebra basis as \[ \varpi(\mathrm{d}n)=\left[ \begin{array}{ccc} 0 & 0 & 0 \\ 0 & 1 & 0 \\0 & 0 & 0 \\\end{array}\right],\qquad \varpi(\mathrm{d}t)=\left[ \begin{array}{ccc} 0 & 0 & 1 \\ 0 & 0 & 0 \\ 0 & 0 & 0 \\\end{array}\right],\]and thus the pseudo-counting increment $\mathrm{d}\mathbf{n}$ is dilated to $\pi(\mathrm{d}t)\otimes\varpi(\mathrm{d}n)$. The reason why we have denoted the $\star$-algebra representation by $\varpi$ is so not to confuse it with the representation $\pi$ of the nilpotent algebra introduced in previous chapters, which requires only $2\times2$ matrix algebra.
\subsection{The Output Martingale}
For the purposes of this discussion a martingale may be regarded as a process of independent increments with zero expectation. In the context of the dynamics (\ref{21}) we must establish the \emph{output measure} to which the expectation of an {output martingale} is defined. This is generally given by the dilated interaction operator
\[
\mathscr{G}_\lambda=\left[
  \begin{array}{ccc}
    I & 0 & -\mathrm{i}\lambda H \\
    0 & \mathbf{G}^\lambda & 0 \\
    0 & 0 & I \\
  \end{array}
\right]\equiv\mathscr{I}+\mathscr{L}_\lambda,\quad \mathscr{L}_\lambda=\frac{\mathrm{i}}{\lambda}H \varpi(\mathrm{d}\mathbf{n})-\mathrm{i}\lambda H\varpi(\mathrm{d}t)
\]
where $\mathscr{I}$ is the identity matrix. With respect to this interaction operator the dilation of (\ref{21}) is a pure counting process
\begin{equation}
\mathrm{d}\Psi^\nu_t(\vartheta)=\mathscr{L}_\lambda\Psi^\nu_t(\vartheta)\mathrm{d}n_t(\vartheta)\label{23}
\end{equation}
which may be compared to (\ref{6}). Here $\Psi^\nu_0$ is simply an embedding of $\psi^\nu_0$ by $\boldsymbol{F}^\star$ %the embedding operator $\boldsymbol{\mathscr{F}}^\star$ corresponding to the composition of $\psi_0$ with the second quantization of an input gauge vector $\chi_+=0\oplus0\oplus1$ having an additional degree of freedom. Indeed
and $\psi^\nu_t=\boldsymbol{F}\Psi^\nu_t$.
\begin{remark}Notice that the generator $\mathscr{G}_\lambda$ may also be given in the form
\[
\mathscr{G}_\lambda=\left[
  \begin{array}{cc}
    \mathbf{I} & \boldsymbol{\upsilon}_\lambda^\star\mathbf{E} \boldsymbol{\upsilon}_\lambda\\
    0 & \mathbf{I} \\
  \end{array}
\right],\quad \mathbf{E}=\left[
  \begin{array}{cc}
    0 & -\mathrm{i}H \\
    \mathrm{i}H & 0 \\
  \end{array}
\right]
\]
so that (\ref{23}) may be obtained directly from (\ref{20}) as a dilation with respect to an extended apparatus having, for example, the input gauge vector $0\oplus0\oplus1\oplus1$ in $\big(L^1\otimes\mathbb{C}^2\big)\oplus\big(L^\infty\otimes\mathbb{C}^2\big)$ with metric $\boldsymbol{\eta}\otimes\boldsymbol{\eta}$.
\end{remark}
The output measure of the pseudo-counting increment is defined as
\begin{equation}
\widetilde{\texttt{M}}_t\big[\mathrm{d}\mathbf{n}\big]={\psi_t^\nu}^\star\boldsymbol{\xi}_\nu^\star \mathscr{G}_\lambda^\star\varpi(\mathrm{d}\mathbf{n})\mathscr{G}_\lambda\boldsymbol{\xi}_\nu\psi^\nu_t = \nu\mathrm{d}t
\end{equation}
where $\boldsymbol{\xi}^\star_\nu=(1,(\sqrt{\nu},0),0)$ and we may choose the increment
\[
\mathrm{d}\widetilde{\boldsymbol{m}}=\frac{1}{\lambda}\big(\mathrm{d}\mathbf{n}-\nu\mathrm{d}t\big)
\]
to define the output martingale $\widetilde{\boldsymbol{m}}$. Now the partially dilated pseudo-measurement process may be described as
\begin{equation}
\mathrm{d}\psi_t^\nu(\vartheta)=\mathrm{i}H\psi_t^\nu(\vartheta) \mathrm{d}\widetilde{\boldsymbol{m}}_t(\vartheta)-\mathrm{i}\Big(\lambda- \frac{\nu}{\lambda}\Big)H\psi_t^\nu(\vartheta)\mathrm{d}t
\end{equation}
having the same deterministic propagation as (\ref{20}) but now with an additional stochastic term.

The next thing we shall do is set $\lambda=\sqrt{\nu}$. This renders the Schr\"odinger equation (\ref{20}) trivial as it becomes
\[
\mathrm{d}\psi(t)=0
\]
so that there is a static wave-function in the ignorance of the measurement process, but we still have two non-trivial dynamical equations. The first is that we still have a free evolution of an unobserved quantum system given by (\ref{25}) with $\lambda=\sqrt{\nu}$, and the second is that we still have the non-trivial pseudo-measurement process of an otherwise freely evolving quantum system. This is given by
\begin{equation}
\mathrm{d}\psi_t^\nu(\vartheta)=\mathrm{i}H\psi_t^\nu(\vartheta) \mathrm{d}\widetilde{\boldsymbol{m}}^\nu_t(\vartheta)\label{26}
\end{equation}
where
\[
\mathrm{d}\widetilde{\boldsymbol{m}}^\nu_t=\frac{1}{\sqrt{\nu}}\mathrm{d}\mathbf{n}_t-\sqrt{\nu}\mathrm{d}t.
\]

The central-limit for the output of this pseudo-measurement process is easily obtained by transforming the increment representation  $\varpi(\mathrm{d}\widetilde{\boldsymbol{m}}^\nu_t)$ into the Fock-vacuum basis
\[
\varpi(\mathrm{d}\widetilde{\boldsymbol{m}}^\nu_t)\mapsto\mathscr{Z}_\nu^\star \varpi(\mathrm{d}\widetilde{\boldsymbol{m}}^\nu_t)\mathscr{Z}_\nu:= \varpi(\mathrm{d}\widetilde{\boldsymbol{\zeta}}^\nu_t),
\]
as was done by Belavkin in the standard case, see \cite{Be00b}, where $\mathscr{Z}_\nu$ is the $\star$-unitary operator (\ref{65}) generating the $\star$-Weyl transform $\mathbf{W}_\nu=\boldsymbol{F}\mathscr{Z}_\nu^\otimes\boldsymbol{F}^\star$ that maps the Fock-vacuum $\delta_\emptyset$ to the Poisson embedding $\boldsymbol{F}^\star_\nu$. Under this transformation equation (\ref{26}) becomes
 \begin{equation}
\mathrm{d}\psi_t^\emptyset(\vartheta)=\mathrm{i}H\big[\psi_t^\emptyset \mathrm{d}\widetilde{\boldsymbol{\zeta}}^\nu_t\big](\vartheta),\qquad \psi_0^\emptyset:=\psi\otimes0^\otimes\label{27}
\end{equation}
 and in the central limit $\sqrt{\nu}\nearrow\infty$ we obtain the pseudo-Wiener process $\widetilde{\boldsymbol{w}}^t_-$ given with respect to the output gauge vectors $\xi_-$ as
 \[
 \lim_{\sqrt{\nu}\nearrow\infty} \mathrm{d}\widetilde{\boldsymbol{\zeta}}^\nu_t= \mathrm{d}a_t\otimes\xi_-^\star(t)+\mathrm{d}a_t^\star\otimes\xi_-(t):=\mathrm{d}\widetilde{\boldsymbol{w}}^t_-.
 \]
 The differential increments $\mathrm{d}a$ and $\mathrm{d}a^\star$ are the remaining two fundamental increments of quantum stochastic calculus. Respectively they are the increments of annihilation and creation, represented as
 \[
\varpi(\mathrm{d}a)=\left[
  \begin{array}{ccc}
    0 & 1 & 0 \\
    0 & 0 & 0 \\
    0 & 0 & 0 \\
  \end{array}
\right],\qquad \varpi(\mathrm{d}a^\star)=\left[
  \begin{array}{ccc}
    0 & 0 & 0 \\
    0 & 0 & 1 \\
    0 & 0 & 0 \\
  \end{array}
\right],
\]
 and here the processes of creation and annihilation are defined on the Fock space of the pseudo-measurement process. Notice that the $\star$-Weyl transform generated by $\mathscr{Z}$ introduces these two remaining $\star$-algebra basis elements into the mechanics.

 The boosted time-increment of the free-evolution of the quantum system diverges in the central limit of the measurement process, but this does not cause a problem as it is no longer meaningful to consider such free-evolution when the object becomes under continuous observation.

 Now the measurement process of the quantum system becomes a pseudo-diffusive output process in the vacuum:
 \[
 \mathrm{d}\psi_t^\emptyset(\vartheta)=\mathrm{i}H\big[\psi_t^\emptyset(\vartheta) \mathrm{d}\widetilde{\boldsymbol{w}}^t_-\big](\vartheta).
 \]
 However, this quantum pseudo-Wiener process is nilpotent and its representation coincides with that of the deterministic increment such that $\varpi(\mathrm{d}\widetilde{\boldsymbol{w}}^t_-)=\pi(\mathrm{d}t)\otimes\mathbf{I}$, only the expectation obtained by tracing over the apparatus indeed yields no dynamics $\mathrm{d}\psi(t)=0$ as the input vector is of the form $\xi_+\otimes\xi_+$.

%Pseudo-measurement may be used to verify a stationary function $\mathrm{d}\psi=0$. We assume that the measurements are $\star$-unitary and it follows that system has a free dynamics in the absence of measurement, but in the ignorance of measurement is does not evolve. In the central limit of the sequential measurement the free dynamics of the system diverges, but what is observed is the creation of output $\xi_-$ whilst the quantum object transforms $\psi\mapsto-\mathrm{i}H\psi$ each time an output is created. So the free dynamics of the object is no more, and its divergence is no longer of concern.

\section{Conclusion}
It is very nice to be able to derive a pseudo-measurement dynamics from the Schr\"odinger equation, and what we have done to achieve this may also be regarded as a \emph{pseudo}-Stinespring dilation. The author would like to emphasize that it was Belavkin who discovered the matrix-algebra formulation of differential increments; here we simply review some of the consequences that follow in the deterministic case studied more extensively by the author in \cite{MFB11}.

To finish it shall be interesting to make some remarks about the consequences of this dilation of Hamiltonian dynamics. The first thing to notice is that the differential Schr\"odinger equation
\[
\mathrm{d}\psi(t)=-\mathrm{i}H\psi(t)\mathrm{d}t
\]
describes the deterministic dynamics of a closed system. One might then argue the meaning of considering such a closed system. Either it is considered from the inside or the outside. If it considered from the inside then all kinds of problems are raised. The first of these is: what is inside the system doing the considering? This is resolved somewhat by a decomposition of the system into a subsystem and its compliment. The subsystem may now be regarded as an observing system that is extracting data from the compliment. This imposes much additional structure on the system. On the other hand, the system may be considered from the outside. Now we have the question: what is outside the system doing the considering? In this case we have to compose the system with an observer of some kind.

However, from understanding that $\mathrm{d}t$ may be represented, precisely, as a matrix we may `open up' the evolving system by separating its evolution into an un-evolving system and time; this \emph{time} is characterized by a massless wave-function having \emph{past} and \emph{future} degrees of freedom. Only in this way may the closed-system Schr\"odinger dynamics be realized as a marginal dynamics of a greater observation-based process. This has a very deep significance because it means that closed systems may be inferred as the result of a greater interaction dynamics.
\\
\linebreak
This dilation of quantum Hamiltonian dynamics admits a statement of the form: the evolution of a closed system may be understood as a consequence of the observation of this system made by an apparatus.

Whatever this `apparatus' is, its observations' generate time, and its existence may be derived even in the case of general non-adapted quantum stochastic calculus. In other words, the matrix-algebra representation of a general quantum stochastic evolution reveals all increments of evolution as the result of interaction of a system with a `time-generating' apparatus.
One may always refer to an apparatus as an observer: apparatus observes. Here we simply establish that dynamics may be regarded as generated by a canonical observer.

\end{document}